\documentclass[prb,twocolumn,showpacs,floatfix,amsmath,amssymb] {revtex4}
\usepackage{graphicx}

\begin{document}

\title{Low-energy excess of vibrational states in v-SiO$_2$: the role of
transverse dynamics}

\author{O. Pilla$^1$, A. Fontana$^1$, J. R. Gon\c{c}alves$^{1,2}$, G. Viliani$^1$, L. Angelani$^3$}

\affiliation{$^1$INFM and Dipartimento di Fisica, Universit\`a di Trento, 38050 Povo,
Trento, Italy \\
             $^2$Departamento de F\'{\i}sica, Universidade Federal do Cear\'a, C.P. 6030, 
Campus do Pici, 60455-760 Fortaleza, Cear\'a, Brazil \\ 
$^3$INFM, SMC-INFM, and Dipartimento di Fisica, Universit\`a di Roma La Sapienza, 
Piazzale Aldo Moro 2, 00185 Roma, Italy}

\date{\today}

\begin{abstract}
The study of the effects of the density variations on the vibrational dynamics in vitreous silica
is presented. A detailed analysis of the dynamical structure factor, as well as of the current spectra,
allows the identification of a flattened transverse branch which is highly sensitive to the density
variations. The experimental variations on the intensity and position of the Boson Peak (BP) in v-SiO$_2$ 
as a function of density are reproduced and interpreted as being due to the shift and disappearance 
of the latter band. The BP itself is found to correspond to the lower energy tail of the excess states due 
to the piling up of vibrational modes at energies corresponding to the flattening of the transverse branch.
\end{abstract}


\maketitle

\section{Introduction}
In this paper we address one of the still open questions in physics of glasses:
the origin of the excess of vibrational states in the low energy region with respect
to the usual Debye density of states.~\cite{Buchenau:86,Fontana:02} This excess, 
characteristic of every glassy system, gives rise to the presence, in the region below 
a few meV, of a more or less pronounced peak, observed both in Inelastic X-ray-Scattering 
(IXS), Inelastic Neutron-Scattering (INS), and Raman spectra. This peak is universally 
referred to as Boson Peak (BP). Despite its universality, different interpretations 
and hypotheses have up to now been proposed both on its origin and on the nature 
of the excess vibrational states. On the basis of experimental observations and computer
simulations, these modes have been catalogued as spatially localized,~\cite{Buchenau:86,Foret:96} 
spatially  delocalized and propagating,~\cite{Benassi:96,DellAnna:98,Pilla:00} spatially 
delocalized but diffusive in character.~\cite{Feldman:99} Theoretical investigations, 
on the other hand, have interpreted the BP in glasses as arising from the lowest-energy 
van Hove singularity of the corresponding crystal;~\cite{Taraskin:01,Taraskin:02} or  
as the precursor of a dynamical instability expected in a disordered structure;~\cite{Grigera:01} 
or as arising from the high frequency dynamics of a model glass within a \mbox{mode-coupling-like} 
description of G\"otze~\cite{Gotze:00} and of~Theenhaus.~\cite{Theenhaus:01}

Particularly in the case of v-SiO$_2$ at normal density the most widely accepted
explanation is that the BP originates from the piling up of modes near the first 
van-Hove singularity of the transverse acoustic vibrational 
branch.~\cite{Taraskin:01,Taraskin:02} These modes should mainly involve relative rotations 
of almost rigid SiO$_4$ tetrahedra as early pointed out by Buchenau~and~coworkers.~\cite{Buchenau:86} 
In addition  to the ambient pressure data, a much useful information is available in the case of 
v-SiO$_2$ for higher densities, obtained both  from  {\it in-situ} measurements and from permanently 
densified samples. In general, densification results in a shift of the BP towards higher energies and 
in a simultaneous decrease of its intensity.~\cite{Sugai:66,Zha:94,Inamura:99,Inamura:00} 
In addition to the experimental data, the BP behavior with density has been successfully reproduced by 
a series of \mbox{simulations.~\cite{Lacks:00,Jund:00,Pilla:03a}} Recently, very accurate INS measurements on 
densified vitreous silica (d-SiO$_2$) have confirmed these effects.~\cite{Foret:02,Courtens:02,Ruffle:03} 
In densified samples the BP lies at higher energies than in vitreous silica at normal density and 
has a rather weaker intensity. In densified samples the inelastic signal in the BP energy range 
shows a more marked dependence on the scattering wavevector $Q$ as regard its intensity 
and  peak position.  We present here simulation results on v-SiO$_2$ at different densities which, 
when compared with the existing experimental data, can help in clarifying the origin and the 
nature of the excess modes as well as the intensity and shift effects on the BP as a function of density.

\section {Simulation}

The simulated system was a glass consisting of 680 SiO$_2$ molecules interacting 
through the modified BKS two-body interaction potential,~\cite{vanBeest:90,Guissani:96} with 
periodic boundary conditions. After a long thermalization at a density of $2.2$~g~cm$^{-3}$ and
at 5000 K the system was slowly cooled down to 300 K at constant volume. At this temperature, an energy 
minimization was performed to locate the equilibrium configuration at room pressure. 
The box size was then stepwise scaled by 1.5\% amounts up to the final density of 
$4.0$ g cm$^{-3}$. Once reached this value, the density was decreased in the same way.  
At each step the energies of the individual realizations were minimized using geometrical methods
to obtain the relative equilibrium configurations. This procedure mimics a real hydrostatic 
pressure experiment at room temperature as described in detail elsewhere.~\cite{Pilla:03a} 
The pressure was estimated by using the virial theorem and compared to experimental 
data.~\cite{Zha:94,Pilla:03b} After the density cycle the zero pressure realization had a 
density of about $2.8$ g cm$^{-3}$. The dynamical properties of the two zero pressure 
configurations, and of the higher density ones, were computed in the harmonic approximation 
by diagonalizing  the dynamical matrix. In particular, we have computed the density of states
$g(E)$ and the longitudinal and transverse currents spectra ($C(Q,E)$) which, in the one-excitation 
approximation, are given by:

\begin{eqnarray}
\label{corrl}
\nonumber
C^{L}_{_{\alpha\beta}}(Q,E) = & \frac{K_BT}{\sqrt{M_\alpha M_\beta}}
 \cdot \Sigma_p |\Sigma_n (\hat Q \cdot \bar e_p(n)) \exp{{(i \ \bar Q
\cdot \bar R_n )} |^2} \\ & \nonumber \cdot \, \delta(E-E_p) \\
\label{corrt}
\nonumber
C^{T}_{_{\alpha\beta}}(Q,E) = & \frac{K_BT}{\sqrt{M_\alpha M_\beta}}
 \cdot \Sigma_p |\Sigma_n (\hat Q \times \bar e_p(n)) \exp{{(i \ \bar Q
\cdot \bar R_n )} |^2} \\ & \cdot \, \delta(E-E_p)
\end{eqnarray}
where $\alpha,\beta$ indicate Si and O, and $\hat Q$=$\bar Q/|Q|$.

The current spectra in transverse polarization are, contrary to the longitudinal currents, not 
experimentally measurable, and thus they are not directly observable by means of inelastic scattering 
experiments at least in the low-$Q$ and low-$E$ ranges. They are nonetheless very useful in 
understanding the details of the vibrational dynamics, as will be explained in the following section.

\section {Discussion}
\subsection {Current Spectra}

\begin{figure}
\begin{center} 
\includegraphics[keepaspectratio,width=1.08\columnwidth]{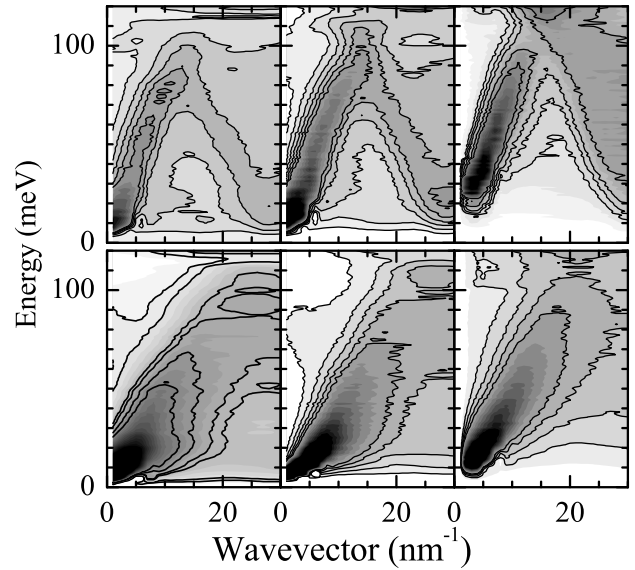} 
\end{center} 
\caption{\label{fig:figure1} 
Computed currents relative to neutron scattering grey scale maps for \mbox{v-SiO$_2$} at three 
different densities (from left to right): \mbox{$\rho=2.2$~g~cm$^{-3}$ ($P=0$~GPa)}, 
\mbox{$\rho=2.8$~g~cm$^{-3}$ ($P=0$~GPa} densified); \mbox{$\rho=4.0$~g~cm$^{-3}$ ($P=35$~GPa)}. 
Upper panels: longitudinal currents; lower: transverse.
The current spectra have been divided by $Q^2$ for visualization purposes.
} 
\end{figure}

Usually the dynamical data obtained via simulation are reported showing the dynamical structure factor, 
$S(Q,E)$. This is mainly because this procedure has the advantage of being directly comparable to the 
experimental spectra obtained, for instance, from standard INS and IXS measurements. On the other hand, 
the current spectra, defined as $C(Q,E)=E^2 S(Q,E)/Q^2$, are much more useful when one is interested in 
characterizing the vibrational spectra. In fact, each current spectrum, taken at a fixed $E$ and $C(Q,E)$, 
gives the $Q$-characterization of the individual vibration with energy $E$, being the current strictly 
related (in harmonic approximation) to the power spectrum of the Fourier transform of the corresponding 
eigenvector.  The computed current spectra, both for longitudinal and transverse dynamics, are shown in 
Fig.~\ref{fig:figure1}(upper and lower panel, respectively) as a contour map plot for v-SiO$_2$ at 
$\rho=2.2, \,\,2.8,\,\, 4.0 $ g cm$^{-3}$. 
The most evident feature  is that in the longitudinal  currents a periodicity is present, even 
if much less evident than in a crystal. The same feature is present in the transverse ones, but with a longer 
quasi period which extends beyond the limit of the figure.

As discussed elsewhere,~\cite{Taraskin:98,Pilla:03a} this difference is due to the fact that the 
longitudinal dynamics is affected  by the average height of the SiO$_4$ tetrahedron, which is 
the main structural unit of v-SiO$_2$, while the transverse vibrations are sensible to the 
average height of SiO$_4$ tetrahedron decorated with oxygen atoms. In the low-$Q$ longitudinal 
spectrum, in addition to the quasi-periodic pattern, two well observable features are present
in the two lowest-density samples, which consist of protunding  peninsulae, nearly parallel 
to the $Q$ axis, centered at about $20$ and $100$  meV. These features have been 
interpreted~\cite{Taraskin:98} as a trace of the vibrational dynamics of the crystobalite 
structure (the crystalline counterpart of v-SiO$_2$), which shows in that energy range flat 
dispersion bands. It should be noted that the plots of Fig.~\ref{fig:figure1} are shown as a 
function of the modulus of the wavevector, which is a relevant quantity in glasses, thus loosing 
every information on the $Q$-direction. If this convention were used also in representing the 
dispersion curves of a crystal, different $Q$-directions in the Brillouin zone would superimpose 
giving qualitatively the same effect as in Fig.~\ref{fig:figure1}. In the $Q$-$E$ range between 
$5$ and $15$ nm$^{-1}$ and $10$ to $25$ meV (where the flat band is observed  in the longitudinal 
currents) a much more evident flattening is present in the transverse currents.  This coincidence 
has suggested the interpretation that the flattened band in the longitudinal currents arises from 
a spilling of transverse dynamics into the longitudinal one, due to disorder.
~\cite{Leadbetter:69,Pilla:03a,Pilla:03b,Taraskin:98} 

\begin{figure}
\begin{center} 
\includegraphics[keepaspectratio,width=1.0\columnwidth]{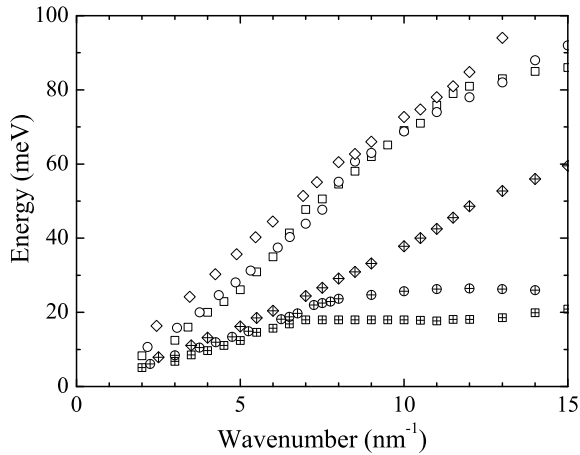} 
\end{center} 
\caption{\label{fig:figure2} 
Main maxima of the longitudinal (open symbols) and transverse (crossed symbols) current spectra, 
for three different densities. Squares: $\rho=2.2$~g~cm$^{-3}$ ($P=0$~GPa); diamonds:  
\mbox{$\rho=4.0$~g~cm$^{-3}$}; circles $\rho=2.8$~g~cm$^{-3}$ ($P=0$~GPa~densified).
} 
\end{figure}

The maxima of the longitudinal and of the transverse currents in the constant $Q$ cuts, are 
reported in Fig.~\ref{fig:figure2} for three different densities. 

In the starting configuration, $\rho=2.2$~g~cm$^{-3}$,  open and crossed squares of 
Fig.~\ref{fig:figure2}, the main maxima relative to the longitudinal currents have a rather defined 
dispersive character up to very high energies (about $100$ meV), while the transverse dispersion one 
flattens at small $Q$-values. The transverse maxima remain centered, within the uncertainty, at the 
same energy where a minor peak is observed also in the longitudinal current spectra, whose intensity 
is roughly 50 to 80\% of the main feature.~\cite{Pilla:03b} By increasing the density up to 
$4.0$~g~cm$^{-3}$, the local structure of the glass, which at normal density consists of fourfold 
coordinated Si ions, is almost completely substituted by an octahedral one (typical of stishovite, 
the crystalline form of silica, stable at high densities). As a consequence, also the dynamical properties 
change and this  can be seen in Fig.~\ref{fig:figure2}. Both longitudinal and transverse maxima 
follow a nearly linear dispersion law up to the highest energies and no flattening is observed. 

\subsection{Boson Peak} 
\begin{figure}
\begin{center}
\includegraphics[keepaspectratio,width=1.0\columnwidth]{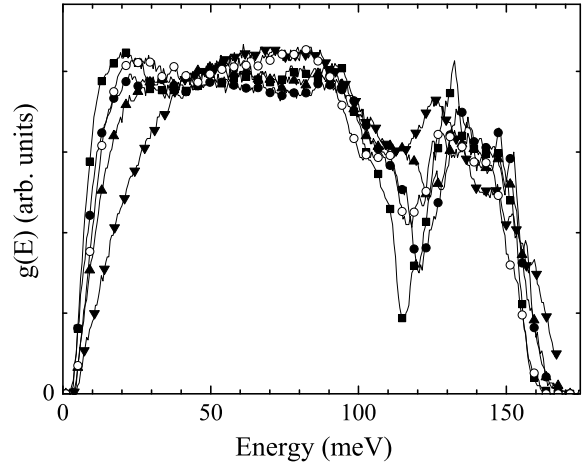} 
\end{center} 
\caption{\label{fig:figure3} 
Vibrational density of states of v-SiO$_2$ at different densities: full squares: $\rho=2.2$~g~cm$^{-3}$; 
full circles: $\rho=2.8$~g~cm$^{-3}$; up triangles: \mbox{$\rho=3.2$~g~cm$^{-3}$;} down triangles: 
$\rho=4.0$~g~cm$^{-3}$; open circles: compressed sample $\rho=2.8$~g~cm$^{-3}$ (zero pressure after
pressure cycle). 
} 
\end{figure}

From our simulations we calculated also the density of states as a function of the density. The result
are shown in Fig.~\ref{fig:figure3} for selected densities. The low energy part of the total density 
of states shifts toward higher frequencies and decreases in intensity  with increasing density. 
As a consequence, the BP  strongly decreases in intensity  and shifts at high energies~\cite{Pilla:03a}. 

\begin{figure}[htb]
\begin{center} 
\includegraphics[keepaspectratio,width=1.0\columnwidth]{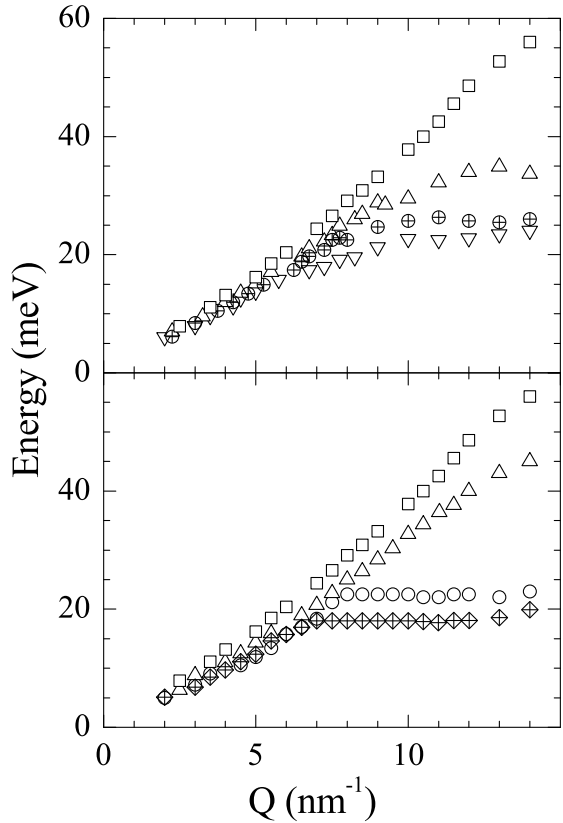} 
\end{center} 
\caption{\label{fig:figure4} Lower panel. Maximum of the transverse current spectra computed at different 
densities with increasing density. Diamonds, circles, triangles and squares correspond 
to $\rho=2.2$~g~cm$^{-3}$, $\rho=2.8$~g~cm$^{-3}$, $\rho=3.2$~g~cm$^{-3}$, and $\rho=4.0$~g~cm$^{-3}$, 
respectively. Upper panel. Maximum of the transverse current spectra measured at different densities 
with decreasing density. Down triangles, circles, up triangles and squares correspond to 
$\rho=2.4$~g~cm$^{-3}$, $\rho=2.8$~g~cm$^{-3}$, $\rho=3.2$~g~cm$^{-3}$, and $\rho=4.0$~g~cm$^{-3}$, 
respectively. The crossed symbols refer to the zero pressure realization, before and after the 
pressure cycle. 
} 
\end{figure}

The reduction of the intensity of the BP with the density can be ascribed to two concurring
effects. The first one is an overall reduction of excess states, induced by the pressure,  and 
the second one is  the  shift towards high energies of the excess which, being weighed by an 
$1/E^2$ factor, appears as having a lower intensity. Actually, both mechanisms are effective in 
\mbox{v-SiO$_2$} as we will discuss later. In Fig.~\ref{fig:figure4} we present the maxima of 
the transverse currents computed at several densities. In the $\rho=2.2$~g~cm$^{-3}$ sample, after 
a linear increase a flattening of the maxima of the transverse currents is observed for $q$-values 
higher than \mbox{$6$ nm$^{-1}$}, at about 15 meV. This plateau progressively shifts, and eventually disappears 
with increasing density. In the highest density sample it is no more observed and only a linear 
dispersion relationship is present. By decreasing the density the plateau is at least in part 
recovered, as shown in the upper panel of  Fig.~\ref{fig:figure4}.  It is worth noting that in the 
densified sample at $P=0$ GPa ($\rho=2.8$~g~cm$^{-3}$) the flattening is present, but it saturates 
at an energy of about 5 meV higher than in the starting realization, indicating an hysteretic behaviour. 

The shift and the disapperance of the plateau help in interpreting the intensity and the energy effects 
on the BP. The excess of states shifts to higher energies and progressively disappears. The fact that 
these effects take place in an energy region, i.e. 15 meV,  higher than the one where the Boson Peak 
is observed, about 4 meV, is only an apparent contradiction. As a matter of fact, the presence of 
the $1/E^2$ factor severely affects the spectral shape of the excess states by weighting only their 
low-energy tail. Indeed, the true excess of states should be observed  only in the {\it difference} 
between the computed density of states and its Debye approximation. 

The determination of the actual shape of the vibrational excess is not a simple task, since it implies 
the knowledge not only of the density of states, but also of the relative weight of the crystalline one,
or at least its Debye approximation. The density of states can be obtained from inelastic neutron 
scattering data, and also, to a certain extent, from unpolarized Raman data. The Debye density of 
states can be independently estimated via Brillouin light scattering, while heat capacity experiments 
give the proper normalization factors.~\cite{Buchenau:86,Carini:95} 

\begin{figure}[htb]
\begin{center} 
\includegraphics[keepaspectratio,width=1.0\columnwidth]{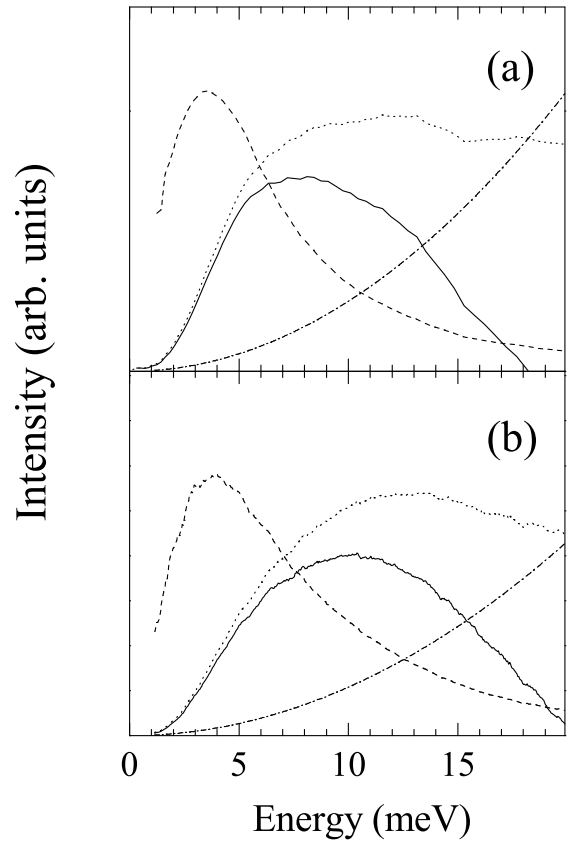} 
\end{center} 
\caption{\label{fig:figure5} 
(a) Dotted line: INS density of states of v-SiO$_2$; dash-dotted-line: Debye approximation; full line: 
vibrational excess, \textit{i.e.} density of states minus Debye; dashed line: density of states divided 
by $E^2$. (b) Same as (a) but for the unpolarized Raman spectrum of v-SiO$_2$. Here the full line represents 
the density of states weighted by the Raman coupling function.}
\end{figure} 

The result of this procedure is shown in Fig.~\ref{fig:figure5}, where INS-deduced $g(E)$ is shown 
with the relative estimate of the Debye contribution. The excess of states (full line in the figure), 
obtained by subtracting the Debye from the DOS, is a very broad unstructured band centered at energies 
higher than $7$ meV. By dividing this latter by $E^2$ its maximum shifts towards lower energies, less 
than $4$ meV, as is shown in Fig.~\ref{fig:figure5} dashed line. This latter peak corresponds to the 
Boson Peak observed both in INS and IXS measurements of the dynamical structure factor. Actually, the 
observed BP lies on the low energy tail of the true excess of states, since the contribution
of the remaining high-energy vibrations is almost suppressed by the $1/E^2$ factor. 
A similar effect is observed also in the depolarized Raman spectra. Here one does not observe 
directly the density of states, but this quantity weighed by the Raman coupling function,~\cite{Fontana:97} 
as shown in Fig.~\ref{fig:figure5}(b). A qualitatively similar behaviour is found even if the scattered 
intensity is somewhat deformed by the Raman coupling function $C(E)$.

The difference between the position of the maxima in the excess of states as found in simulations
(about $17$ meV as in Fig.~\ref{fig:figure4}) and in experiments (about $7$ meV as in Fig.~\ref{fig:figure5}), 
can derive from two main causes. From the computational point of view, the discrepancy may derive from 
the potential used. Moreover, the experimental density of states is obtained through some approximations, 
like f.i. the incoherent approximation and the estimate of the multiphonon contribution. Also, 
in the simulations the excess states could be overestimated due to unphysical quenching times 
currently used in Molecular Dynamics, as pointed out recently by Angell.~\cite{Angell:03}
Finally, the behavior of the transverse current maxima with density shown in Fig.~\ref{fig:figure4}, 
not only accounts for the intensity decrease of the BP, but also explains the coherence effects of 
the BP measured with in INS and IXS as a function of the density.  The BP in normal v-SiO$_2$ has a 
small coherence effect, i.e. it is  rather insensitive to the exchanged wavevector.~\cite{Benassi:96} 
On the contrary, densified samples have a marked coherent behavior at least at small 
$Q$-values.~\cite{Foret:02} From Fig.~\ref{fig:figure4} full circles, one sees that the coherence 
effects in the BP range, i.e. variation in position and/or intensity, are expected in the linear 
dispersing range in the case of the $\rho=2.2$~g~cm$^{-3}$ sample, being the dispersion curve flat 
for wavevectors higher than 6 nm$^{-1}$. In the densified sample at zero pressure, $\rho=2.8$~g~cm$^{-3}$, 
the linear region extends up to 9 nm$^{-1}$. In the latter sample it is indeed possible to study, 
using IXS, a more extended linear region and the direct observation of a Brillouin peak directly 
in the energy cuts is possible in spite of the resolution limitations. Hence, in the densified samples 
our simulations predict (i) in the low-$Q$, low-$E$ range a dispersing region where a Brillouin peak 
is observed extending to wavevectors higher than than in normal silica measurements, (ii) the 
appearance of the BP (nearly $Q$ insensitive) at wavevectors (and corresponding energies) higher 
than in v-SiO$_2$, (iii) a BP intensity weaker than that of the normal density sample. All these 
predictions are confirmed by the experiments of Foret and coworkers,~\cite{Foret:02} while the 
energy shift and the intensity effect have also been reported by Inamura and coworkers.~\cite{Inamura:00}

\section {Conclusions}

In conclusion, the comparison between experiment and simulation allows us to affirm that the BP 
in v-SiO$_2$ corresponds to the low-energy tail of a flattened transverse branch, whose center lies 
at relatively high energies. This is also in agreement with earlier studies,~\cite{Buchenau:86,Taraskin:97} 
which ascribed these vibrations to hindederd  librations of coupled SiO$_4$ tethrahedra. The effect 
responsible for the occurrence of the BP which we have described here is valid for 
v-SiO$_2$, and it cannot be excluded that other mechanisms are active for different systems, like the 
idealized ones studied in Ref.~\onlinecite{Grigera:01}. Nevertheless, it suggests a general approach for 
the study of the BP in glasses. It is well known that every kind of disorder induces a sort of level 
repelling and a consequent piling up of modes at the extremes  of the vibrational spectrum, as shown 
by Elliott and coworkers, (see for instance Ref.~\onlinecite{Simdyankin:02} and references therein). 

\section {Acknowledgments}
We are grateful to Giancarlo Ruocco, Uli Buchenau and Stephen Elliott for very helpful discussions. 
This work was financially supported by INFM by PRAGENFDT, Italian Ministero degli Affari Esteri, and 
MURST Progetto di Ricerca di Interesse Nazionale.


\begin{thebibliography}{99}

\bibitem{Buchenau:86}
   U. Buchenau, M. Prager, N. N\"ucker, A. J. Dianoux, N. Ahmad and W. A. Philips,
   Phys. Rev. B {\bf 34}, 5665 (1986).

\bibitem{Fontana:02}
   A. Fontana and G. Viliani,
   Phil. Mag. B {\bf 82}, (special issue) (2002).

\bibitem{Foret:96}
   M. Foret, E. Courtens, R. Vacher and J. B. Suck,
   Phys. Rev. Lett. {\bf 77}, 3831 (1996).

\bibitem{Benassi:96}
   P. Benassi, M. Krisch, M. Masciovecchio, V. Mazzacurati, G. Monaco, G. Ruocco, F. Sette and R. Verbeni,
   Phys. Rev. Lett. {\bf 77}, 3835 (1996).

\bibitem{DellAnna:98}
   R. Dell'Anna, G. Ruocco, M. Sampoli and G. Viliani,
   Phys. Rev. Lett. {\bf 80}, 1236 (1998).

\bibitem{Pilla:00}
O. Pilla, A. Cunsolo, A. Fontana, C. Masciovecchio, G. Monaco, M. Montagna, G. Ruocco, 
T. Scopigno and F. Sette,
Phys. Rev. Lett. {\bf 85}, 2136 (2000).

\bibitem{Feldman:99}
   J. L. Feldman, P. B. Allen and S. R. Bickham,
   Phys. Rev. B {\bf 59}, 3551 (1999).

\bibitem{Taraskin:01}
   S. N. Taraskin, Y. L. Loh, G. Natarajan and S. R. Elliott,
   Phys. Rev. Lett. {\bf 86}, 1255 (2001).
  
\bibitem{Taraskin:02}
   S. N. Taraskin, J. J. Ludlam, G. Natarajan and S. R. Elliott,
   Phil. Mag. B {\bf 82}, 197 (2002).

\bibitem{Grigera:01}
T. S. Grigera, V. Mart\`in-Mayor, G. Parisi and P. Verrocchio,
Phys. Rev. Lett. {\bf 87}, 085502 (2001).

\bibitem{Gotze:00}
   W. G\"otze and M. R. Mayr,
   Phys. Rev. E {\bf 61}, 587 (2000).

\bibitem{Theenhaus:01}
   T. Theenhaus, R. Schilling, A. Latz and M. Letz,
   cond-mat/0105393 (2001).

\bibitem{Sugai:66}
   S. Sugai and A. Onera,
   Phys. Rev. Lett. {\bf 20}, 4210 (1966).

\bibitem{Zha:94}
   C. S. Zha, R. J. Hemley, H. K. Mao, T. S. Duffy and C. Meade,
   Phys. Rev. B {\bf 50}, 13105 (1994).

\bibitem{Inamura:99}
   Y. Inamura, M. Arai, O. Yamamuro, A. Inaba, N. Kitamura, T. Otomo, T. Matsuo, S. M. Bennington and A. C. Hanon,
   Physica B {\bf 263/264}, 299 (1999).

\bibitem{Inamura:00}
   Y. Inamura, M. Arai, T. Otomo, N. Kitamura and U. Buchenau,
   Physica B {\bf 284-288}, 1157 (2000).
  
\bibitem{Lacks:00}
   D. J. Lacks,
   Phys. Rev. Lett. {\bf 84}, 4629 (2000).

\bibitem{Jund:00}
   P. Jund and R. Jullien,
   J. Chem. Phys. {\bf 113}, 2768 (2000). 

\bibitem{Pilla:03a}
   O. Pilla, L. Angelani, A. Fontana, J. R. Gon\c{c}alves and G. Ruocco,
   J. Phys.: Condens. Matter {\bf 15}, S995 (2003).

\bibitem{Foret:02}
   M. Foret, R. Vacher, E. Courtens and G. Monaco,
   Phys. Rev. B  {\bf 66}, 024204 (2002).

\bibitem{Courtens:02}
   E. Courtens, M. Foret, B. Helen, B. Ruffl\'e and R. Vacher,
   Phys. Rev. B {\bf 66}, 024204 (2002).

\bibitem{Ruffle:03}
   B. Ruffl\'e, M. Foret, E. Courtens, R. Vacher and G. Monaco,
   J. Phys. Condens. Matter {\bf 15}, 1281 (2003).

\bibitem{vanBeest:90}
   B. W. H. van Beest, G. J. Kramer and R. A. van Santen,
   Phys. Rev. Lett. {\bf 64}, 1955 (1990).

\bibitem{Guissani:96}
   Y. Guissani and B. Guillot,
   J. Chem. Phys. {\bf 104}, 7633 (1996).

\bibitem{Pilla:03b}
   O. Pilla, S. Caponi, A. Fontana, M. Montagna, F. Rossi, G. Viliani, L. Angelani, G. Ruocco, G. Monaco and F. Sette,
   cond-mat/0209519 (2003).

\bibitem{Taraskin:98}
   S. N. Taraskin and S. R. Elliott,
   Phil. Mag. B  {\bf 2}, 403 (1998).

\bibitem{Leadbetter:69}
   A. J. Leadbetter,
   J. Chem. Phys. {\bf 51}, 779 (1969).

\bibitem{Carini:95}
   G. Carini, G. D'Angelo, G. Tripodo, A. Fontana, A. Leonardi, G. A. Saunders and A. Brodin,
   Phys. Rev. B {\bf 52}, 9342 (1995-I).

\bibitem{Fontana:97}
   A. Fontana, F. Rossi, G. Carini, G. D'Angelo, G. Tripodo and A. Bartolotta,
   Phys. Rev. Lett. {\bf 78}, 1078 (1997).

\bibitem{Angell:03}
   C. A. Angell, Y. Yue and L-M. Wang, J. R. D. Copley, S. Borik and S. Mossa,
   J. Phys.: Condens. Matter {\bf 15}, S1051 (2003).

\bibitem{Taraskin:97}
   S. N. Taraskin and S. R. Elliott,
   Phys. Rev. B {\bf 56}, 8605 (1997).

\bibitem{Simdyankin:02}
   S. I. Simdyankin, S. N. Taraskin, M. Elenius, S. R. Elliott and M. Dzgutov,
   Phys. Rev. B {\bf 65}, 104302 (2002).

  
\end{thebibliography}
\end{document}